\documentclass[%
reprint,
preprintnumbers,
showkeys,
amsmath,amssymb,
aps,
prl,
floatfix,superscriptaddress,a4paper
]{revtex4-2}

\usepackage[english]{babel}
\usepackage[utf8]{inputenc}
\usepackage[T1]{fontenc}
\usepackage{graphicx}
\usepackage[pdfencoding=auto,psdextra]{hyperref}
\usepackage{xcolor}
\usepackage{amsmath}
\usepackage{amsfonts}
\usepackage{amssymb}

\newcommand{\fd}{\mathop{}\!\mathbin\bigtriangleup}
\newcommand{\ff}[1]{\underline{#1}}

\hypersetup{
    pdftoolbar=true,               
    pdfmenubar=true,               
    pdffitwindow=false,            
    pdfstartview={FitH},           
    pdftitle={},                   
    pdfnewwindow=true,             
    colorlinks=true,               
}

\begin{abstract}
    Creating non-classical states of light from simple quantum systems together with classical resources is a challenging problem.
    We show how chiral emitters under a coherent drive can generate non-classical photon states.
    For our analysis, we select a specific temporal mode in the transmitted light field, resulting in a coupled master equation for the relevant mode and the chiral emitters.
    We characterise the mode's state by its Wigner function and show that the emission from the system predominantly produces mixtures of few-photon-added coherent states.
    We argue that these non-classical states are experimentally accessible and show their application for quantum metrology.
\end{abstract}

\begin{document}

\title{Creation of non-classical states of light in a chiral waveguide}

\author{Kevin Kleinbeck}
\affiliation{Institute for Theoretical Physics III and Center for Integrated Quantum Science and Technology,\\ University of Stuttgart, Pfaffenwaldring 57, 70550 Stuttgart, Germany}
\author{Hannes Busche}
\author{Nina Stiesdal}
\author{Sebastian Hofferberth}
\affiliation{Institute for Applied Physics, University of Bonn, Wegelerstraße 8, 53115 Bonn, Germany}
\author{Klaus Mølmer}
\affiliation{Niels Bohr Institute, University of Copenhagen, Blegdamsvej 17, 2100 Copenhagen, Denmark}
\author{Hans Peter B\"uchler}
\affiliation{Institute for Theoretical Physics III and Center for Integrated Quantum Science and Technology,\\ University of Stuttgart, Pfaffenwaldring 57, 70550 Stuttgart, Germany}

\date{\today}

\maketitle

\section{Introduction}
Non-classical states of light are an essential ingredient not only in optical quantum technology, but also many fundamental physics experiments.
Prominent examples include the use of NOON- and squeezed states in quantum metrology~\cite{afek2010,Abadie2011,aasi2013}, the violation of Bell inequalities using entangled photon pairs~\cite{weihs1998,christensen2013,giustina2015}, or single photons as information carriers in quantum cryptography~\cite{stucki2002,gobby2004,peev2009,lunghi2013} and information processing~\cite{bentivegna2015,larsen2019}.
Meanwhile, the generation of highly non-classical states of light is not straightforward.
For example, squeezed states are limited to few decibel~\cite{kim1994,tse2019}, state of the art NOON states are still limited to the few photon regime~\cite{hua2014}, and many protocols to generate non-classical light require heralding or post-selection~\cite{zavatta2004,kim2009}.

In recent years, quantum emitters coupled to chiral waveguides in which light propagates in a single, well-defined direction, have emerged as promising experimental tools for the manipulation of light~\cite{bliokh2015,lodahl2017}.
For example, it has been shown that these systems can be used to implement atom-mediated photon-photon interactions, photon circulators, deterministic photon sources, and single to few photon subtractors~\cite{mitsch2014,sollner2015,scheucher2016,stiesdal2018,stiesdal2021,lu2021}.
They can be implemented in a variety of different platforms, ranging from circuit QED with superconducting qubits~\cite{hoi2012,chapman2017}, to quantum dots coupled to photonic crystal waveguides~\cite{arcari2014,sollner2015,xiao2021}, atoms coupled to optical nanofibres~\cite{mitsch2014,scheucher2016,solano2017}, or free space Rydberg superatoms~\cite{paris2017}.
From a theoretical perspective, chiral waveguides can be considered as directed quantum networks, where the output of each quantum node adds to the input of all subsequent nodes.
Consequently, powerful theoretical tools are available to investigate chiral quantum systems, like input-output relations for the emitted light~\cite{gardiner2004,combes2017} or descriptions by exact Lindblad master equation in the presence of multiple emitters~\cite{pichler2015,shi2015}.

In this work, we propose a simple scheme for the generation of non-classical light by scattering classical light on a cascaded chain of quantum emitters in a chiral waveguide, circumventing the aforementioned problems of post-selection or heralding.
The main idea is that certain light modes of the transmitted light field exhibit highly non-classical character.
In order to study such temporal modes, we describe the chiral waveguide as a quantum input-output network coupled to a virtual photonic cavity~\cite{combes2017,kiilerich2019}, tuned to capture only photons in a specifically selected mode.
For example, this formalism has previously been used to explain experimental results for the steady state emission of a superconductiong qubit~\cite{lu2021}, or the emission of a Rydberg superatom inside an optical cavity~\cite{magro2022}. 
First, we analyse the output for a single emitter, which is non-classical as indicated by negativity in its Wigner function, and we link the temporal evolution of the selected light mode to the Rabi dynamics of the emitter.
Subsequently, we investigate how decoherence and dephasing of the emitter decrease the negativity in the Wigner function.
We then extend our investigation to the generation of non-classical light by scattering on a chain of emitters, where waveguide-mediated emitter-emitter interactions come into play and the formation of bound states of photons influences the number statistics of the observed light mode.
Finally, we outline how the resulting non-classical state becomes accessible in quantum experiments and, as an example, show its application in quantum metrology, where a combination of the non-classical state and a coherent state beats the standard quantum limit of interferometry.

\section{Model}
We consider a chain of $M$ quantum emitters with chiral coupling to the electromagnetic field where each emitter~$i$ has a ground state $|G_i\rangle$, an excited state $|W_i\rangle$, and an additional non-radiating state $|D_i\rangle$, relevant for the study of dephasing effects.
The dynamics of the emitters driven by a coherent input light source is well-described by a Lindblad master equation~\cite{pichler2015,shi2015}.
For the chiral waveguide the transmitted photon field is simply a combination of the input field and the response of the emitters to the coherent drive, following the input-output relations~\cite{gardiner2004} $b(t) = \alpha(t) + \sqrt{\kappa} \sigma_\mathrm{chain}^-(t)$.
Here $\alpha(t)$ denotes the amplitude of the incoming light field, $b^{(\dagger)}(t)$ destroys (creates) a photon immediately behind the emitter chain, $\sigma_\mathrm{chain}^- = \sum_{i=1}^M \sigma_i^- = \sum_{i=1}^M |G_i\rangle\langle W_i|$ is the collective decay operator of the chain, and $\sqrt{\kappa}$ denotes the collective coupling strength of the emitters to the photons.


\begin{figure}[t]
    \centering
    \includegraphics[width=\columnwidth]{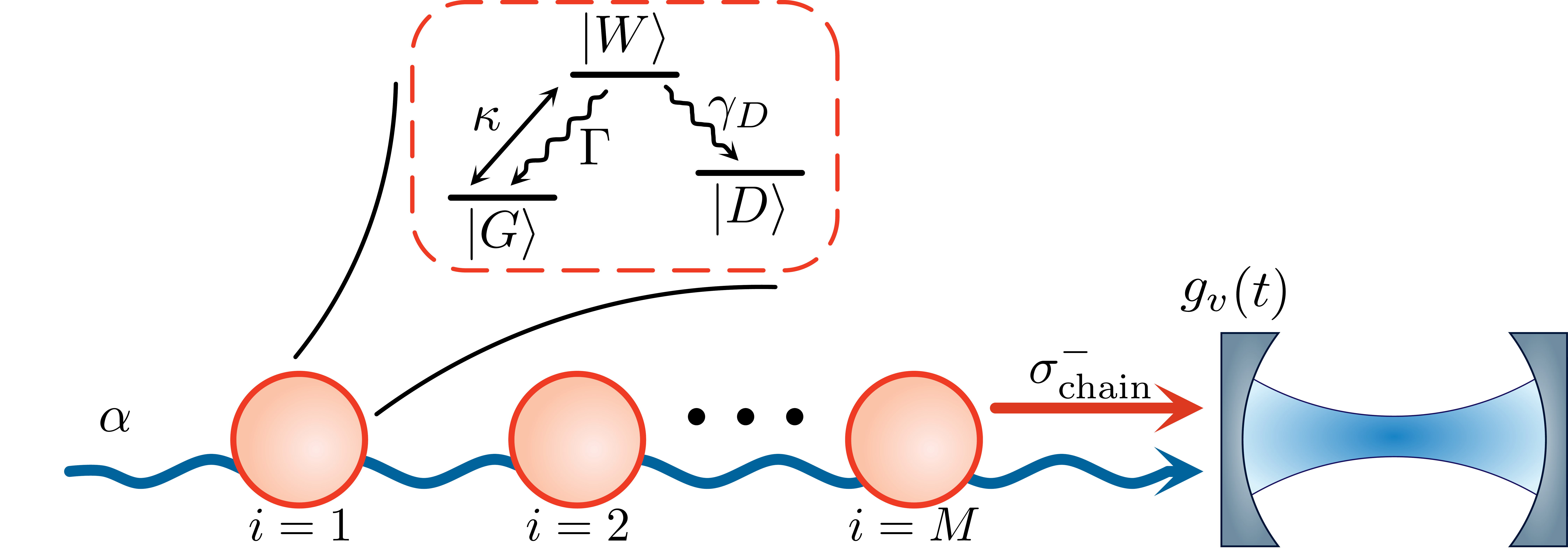}
    \caption{
        Scattering coherent light with amplitude $\alpha$ on a chiral emitter chain produces light in multiple orthogonal modes.
        The light in a specific mode $v(t)$ is studied by placing a virtual cavity behind the emitters, with a time-dependent coupling rate $g_v(t)$ tuned to only capture photons in mode $v(t)$.
        The coherent light drives both the cavity and the $|G\rangle \leftrightarrow |W\rangle$ transition of the emitters.
        Our theory includes emission of $|W\rangle$ out of the waveguide at rate $\Gamma$ and an induced transfer with rate $\gamma_D$ of $|W\rangle$ into a non-radiating excited state $|D\rangle$.
    }
    \label{fig:sketch}
\end{figure}

This work specifically investigates the quantum statistics of a temporal mode $v(t)$ in the transmitted light field.
More specifically, we analyse the occupation in the mode
\begin{equation}
    b_v = \int dt\, v(t) b(t),
\end{equation}
with $v(t) = \frac{1}{\sqrt{\tau}} \Theta(t_0 < t < t_0 + \tau)$.
In the Lindblad master equation formulation the occupation of mode $v(t)$ can be obtained by placing a virtual cavity with the time-dependent coupling rate
\begin{equation}
    g_v(t) = - \frac{v^*(t)}{\sqrt{\int_0^t \mathrm{d}t\, |v(t)|^2}},
\end{equation}
behind the emitters~\cite{kiilerich2019}, as depicted in Figure~\ref{fig:sketch}.
Due to the choice of $g_v$, the cavity only accumulates photons from mode $v(t)$ such that the asymptotic state of the cavity $\rho_v$ equals the state of the photons in mode $v(t)$.

In this description, the emitter-cavity system evolves according to the master equation
\begin{align}
    \notag
    \frac{\mathrm{d}\rho}{\mathrm{d}t} =
    &-\frac{i}{\hbar} [H_\mathrm{drive} + H_\mathrm{sys} + H_\mathrm{exc}, \rho] \\
    &+ D_L[\rho] + \Gamma \sum_{i=1}^M D_{\sigma_i^-}[\rho]
        + \gamma_D \sum_{i=1}^M D_{|D_i\rangle\langle W_i|}[\rho].
    \label{eq:Master}
\end{align}
The coherent background drives both the emitters and cavity
\begin{align}
    \notag
    H_\mathrm{drive} &= i \hbar\big(\alpha^*(t) L - \alpha(t) L^\dagger \big),
\end{align}
with $L = \sqrt{\kappa}\sigma_\mathrm{chain}^- + g_v^*(t) b_v$, while the emitters interact via chiral exchange of virtual photons,
\begin{align}
    H_\mathrm{sys} = - i\hbar\frac{\kappa}{2} \sum_{i > j}
        \Big( \sigma_i^+ \sigma_j^- - \mathrm{H.c.} \Big).
    \label{eq:Hsys}
\end{align}
The Hamiltonian
\begin{equation}
    H_\mathrm{exc} =
        \frac{i}{2} \hbar \big(\sqrt{\kappa} g_v^*(t) \sigma_\mathrm{chain}^+ b_v
        - \mathrm{H.c.} \big)
\end{equation}
describes the coherent exchange interaction between the emitters and the virtual photon cavity where $b_v^{(\dagger)}$ destroys (creates) a photon in the cavity.

The emitter-cavity system is subject to the collective decay $D_L[\rho] = L\rho L^\dagger - 1/2 \{L^\dagger L, \rho\}$.
In addition, we consider photon losses out of the waveguide, described by a decay $\Gamma$ of the excited states into the respective ground state, and a decay $\gamma_D$ of the excited states $|W_i\rangle$ into the non-radiating dark states $|D_i\rangle$.

\section{Results}
\subsection{Creation of non-classical light}
In the following, we investigate whether the state of light in the output mode is non-classical. We base this classification on the Wigner-phase-space distribution~\cite{wigner1932}
\begin{equation}
    W(\beta = x + ip)
    = \frac{1}{\pi} \int dy\, \langle x+y|\rho_v|x-y\rangle
    e^{-2ipy}.
\end{equation}
While $W$ is normalised, it is generally not positive everywhere and thus cannot be regarded as a classical phase-space probability distribution.
Thus, we take the existence of phase space domains where the Wigner function is negative as an indicator for non-classical states and we measure non-classicality by the total negative part of the Wigner function $W^- = \int d^2\beta\, |\min(0, W)|$.
It should be noted however that squeezed states produce Gaussian Wigner functions, yet they are often considered non-classical~\cite{walls1983}.
The Wigner negativity accounts for a stronger kind of non-classicality, e.g., the nonexistence of an efficient classical description~\cite{mari2012}.
 
\subsubsection{Single Emitter}
\begin{figure}[t]
    \centering
    \includegraphics[width=\columnwidth]{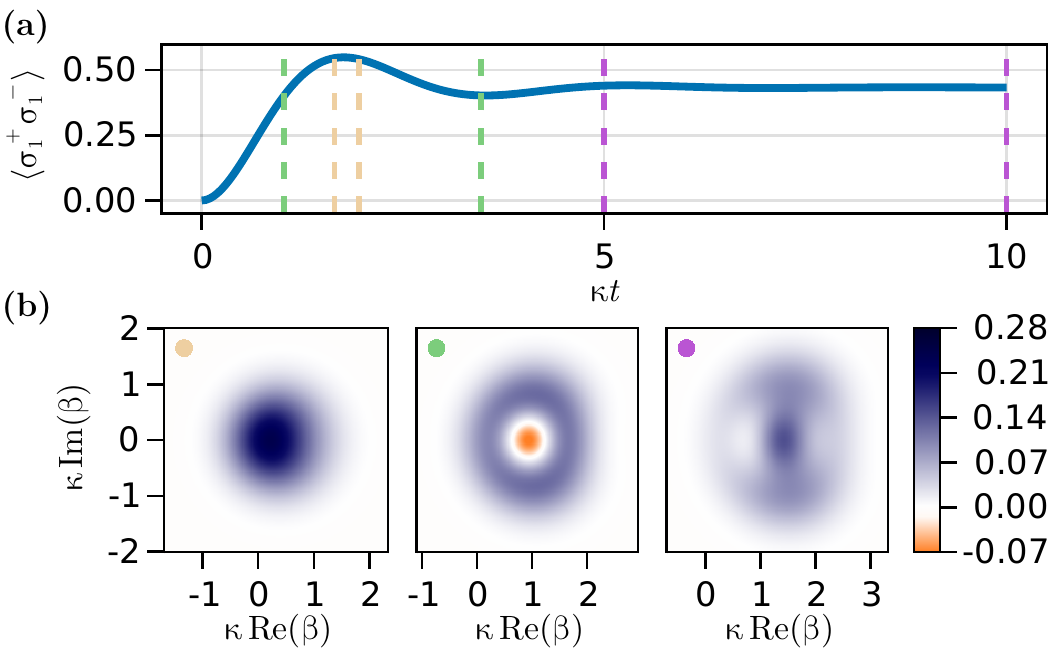}
    \caption{
        \textbf{(a)} A single emitter's excited state population for $\alpha = 0.9\sqrt{\kappa}$, $\gamma_D = 0 = \Gamma$.
        The vertical lines indicate the different choices of binning intervals studied in (b).
        \textbf{(b)} Wigner functions of $\rho_v$ for square mode pulses on the different time intervals.
        For short time bins (left) the Wigner function closely resembles the coherent state $|\sqrt{\tau}\alpha\rangle$, while for $\kappa \tau \gg 1$ the Wigner function becomes positive as the emitter's emission close to its steady state weakly impacts the coherent background (right).
        Bins centered around the first Rabi-peak provide the largest Wigner negativities for suitable $\alpha$ (middle).
    }
    \label{fig:Examples}
\end{figure}

First, we show that for a single emitter, the state of light in mode $v(t)$, $\rho_v = \mathrm{Tr}_\mathrm{emitter}[\rho(t_0 + \tau)]$ possesses quantum mechanical number statistics.
At a given driving strength $\alpha$, the Wigner function of $\rho_v$ primarily depends on the time bin width $\tau$, as displayed in Figure~\ref{fig:Examples}, which shows the time evolution of the excited state population in panel (a).
In the short-binning limit $\kappa \tau \ll 1$, we determine $\rho_v$ analytically with the help of the input-output relations $b_v \approx \sqrt{\tau} (\alpha + \sqrt{\kappa} \sigma_\mathrm{chain}^-(t_0))$~\cite{gardiner2004} resulting in~\cite{Appendix}
\begin{align}
    \notag
    &\mathcal{D}^\dagger(\sqrt{\tau}\alpha) \rho_v \mathcal{D}(\sqrt{\tau}\alpha) = \\
    &\quad
    \Big[
        |0\rangle\langle 0|
        \notag
        + \sqrt{\kappa \tau}
        \big[\langle\sigma_\mathrm{chain}^-(t_0)\rangle |1\rangle\langle 0|
            + \langle\sigma_\mathrm{chain}^+(t_0)\rangle |0\rangle\langle 1|
        \big] \\
        &\quad + \kappa \tau \langle\sigma_\mathrm{chain}^+(t_0)\sigma_\mathrm{chain}^-(t_0)\rangle
        \big(|1\rangle\langle 1| - |0\rangle\langle 0|\big)
    \Big],
    \label{eq:DensityShortBin}
\end{align}
where $\mathcal{D}$ is the displacement operator.
For short $\tau$, $\rho_v$ predominantly resembles a coherent state $|\sqrt{\tau}\alpha\rangle$ with a small admixture of a photon-added coherent state or --- equivalently --- a displaced single-photon state~\cite{zavatta2004}, see Figure~\ref{fig:Examples}(b).
For larger~$\tau$, we can no longer find $\rho_v$ exactly and we solve equation~\eqref{eq:Master} numerically.

We find that bins $(t_0, t_0 + \tau)$ centred around the first peak of the Rabi oscillation are typically optimal for the generation of non-classical light with the presented method, see center plot in Figure~\ref{fig:Examples}(b).
This key observation can be explained by considering the underlying dynamics in two steps.
First, up to time $t = t_0$, the virtual cavity is closed and the emitter is prepared close to its first Rabi peak making subsequent photon emission more likely.
Next, the cavity is opened for a time $\tau$ and absorbs the background photons and photons emitted by the emitter.

The degree to which the light in the cavity is nonclassical depends strongly on $\tau$. If $\tau$ is short compared to the timescale of the Rabi oscillations, the chance of storing additional photons in the cavity becomes negligible.
If $\kappa \tau \gg 1$, then the cavity state will be dominated by the coherent background with some added "noise" due to the emitter signal, and $\rho_v$ generally loses its negative features in the Wigner function in this regime.
The absence of nonclassical character in these cases is evident in the first and third example displayed in Figure~\ref{fig:Examples}(b).
For $\tau$ on the order of the Rabi cycle duration, however, the photon-emitter-interaction has a strong influence on the character of the light in the cavity and the Wigner function exhibits clear negative features. Below, we provide further evidence that $\rho_v$ is well-described by photon-added coherent states.

Without the emitter, the output cavity is only affected by the coherent input and the cavity density matrix becomes $\rho_v(t) = |\tilde{\alpha}(t)\rangle\langle \tilde{\alpha}(t)|$ for the flat mode $v(t)$ and with $\tilde{\alpha}(t) = \alpha/g_v^*(t)$ for $t > t_0$ ($\tilde{\alpha} = 0$ otherwise).
Factoring out the coherent contribution $\rho = \mathcal{D}(\tilde{\alpha}(t))\tilde{\rho}\mathcal{D}^\dagger(\tilde{\alpha}(t))$ shows that the non-displaced part~$\tilde{\rho}$ evolves according to the same master equation~\eqref{eq:Master}, up to the replacement
\begin{equation}
    H_\mathrm{drive} \mapsto i\sqrt{\kappa}(\alpha^*\sigma_\mathrm{chain}^- - \alpha\sigma_\mathrm{chain}^+).
\end{equation}
The time evolution of $\tilde{\rho}$ therefore resembles a system in which a coherently driven emitter may emit its excitations into a non-driven virtual cavity.
As the emitter can only produce temporally separated photons, this explains the photon-added contribution to $\rho_v$.

We find further evidence that $\rho_v$ is a displaced mixture of Fock states, by comparing $\rho_v$ to a generalisation of the exact result~\eqref{eq:DensityShortBin} for $\kappa \tau \ll 1$.
For larger $\tau$, it is expected that the emitter can absorb and re-emit multiple photons within $\tau$ and we thus make the displaced two-photon Ansatz
\begin{equation}
    |\psi_i\rangle = \mathcal{D}(\sqrt{\tau}\alpha)
    \big[a_{i,0}|0\rangle + a_{i,1}|1\rangle + a_{i,2}|2\rangle\big]
    \label{eq:Ansatz}
\end{equation}
and approximate $\rho_v \approx \sum_{i=}^3 p_i |\psi_i\rangle\langle \psi_i|$ by a three state mixture of these candidate states.
Note that this description only provides three free parameters, as $a_{i,j}$ may be chosen real with six of them fixed by orthonormality relations and $p_i$ are fixed by the largest three eigenvalues of $\rho_v$.
For all examples presented in this section, this Ansatz reproduces $\rho_v$ with a fidelity of $> 99\,\%$.
Since the displacement operator is equivalent to translations in phase-space, this reveals the non-classical single-photon and two-photon contributions as the origin of the Wigner negativity.

\begin{figure}[t]
    \centering
    \includegraphics[width=\columnwidth]{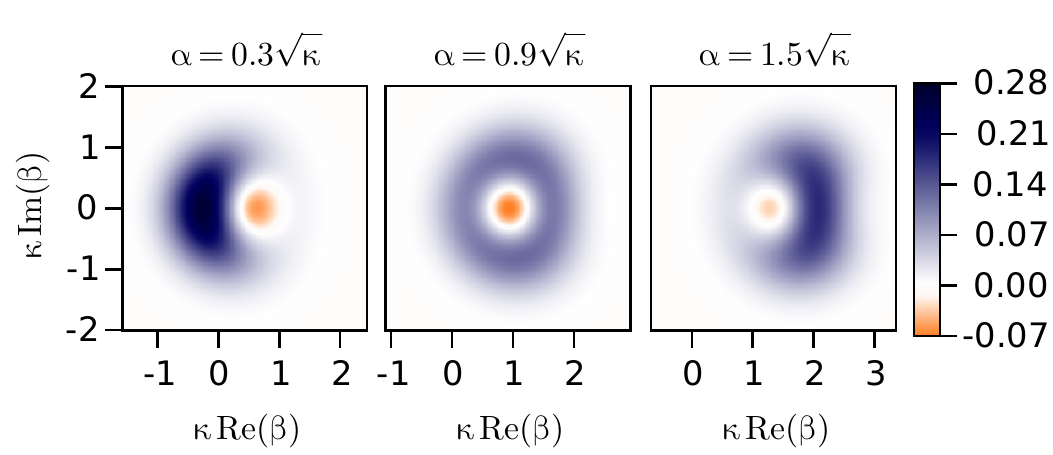}
    \caption{
        Wigner functions of $\rho_v$ at different $\alpha$ and $\gamma_D = 0 = \Gamma$.
        The binning interval at given $\alpha$ is chosen to increase the Wigner negativity at $\gamma_D = 0$ and always includes the first Rabi-peak of the excited state population.
    }
    \label{fig:Wigner}
\end{figure}

Creating non-classical states by binning around the first Rabi peak is possible for driving strengths up to $\alpha \approx 1.5 \sqrt{\kappa}$ as can be seen in Figure~\ref{fig:Wigner}.
For larger $\alpha$, however, binning around the first Rabi peak results again in the $\kappa \tau \ll 1$ limit, discussed above, since the Rabi frequency is $\Omega \approx 2\sqrt{\kappa}\alpha$.
Hence a bin size that only encompasses the first Rabi maximum requires $\tau \propto 1/\sqrt{\kappa}\alpha$, resulting in short bins for large $\alpha$.
As we can see from equation~\eqref{eq:DensityShortBin}, this suppresses the single-photon contribution of $\rho_v$ as $1/\alpha\rightarrow 0$, and $\rho_v$ becomes predominantly coherent.

Up to this point, we have considered a noiseless two-level emitter with perfect chiral emission into the waveguide.
These assumptions are bound to break in any experimental realisation, and we will now discuss the most relevant noise sources and their impact.

First, imperfect chirality results in back-scattering and the emission of photons outside the waveguide.
These two effects result in a spontaneous decay with rate $\Gamma$ for a single emitter.
On the other hand, chiral two-level emitters are commonly mesoscopic, artificial atoms with complex inner structure and dynamics.
For example, a Rydberg superatom consists of a collection of individual atoms, which are subject to thermal motion and intrinsic dipole-dipole interactions.
These effects impact the internal dynamics of the excited state and are well described by an effective decay from the excited state $|W\rangle$ into a non-radiating state $|D\rangle$ with rate $\gamma_D$~\cite{paris2017,stiesdal2018}.

\begin{figure}[t]
    \centering
    \includegraphics[width=\columnwidth]{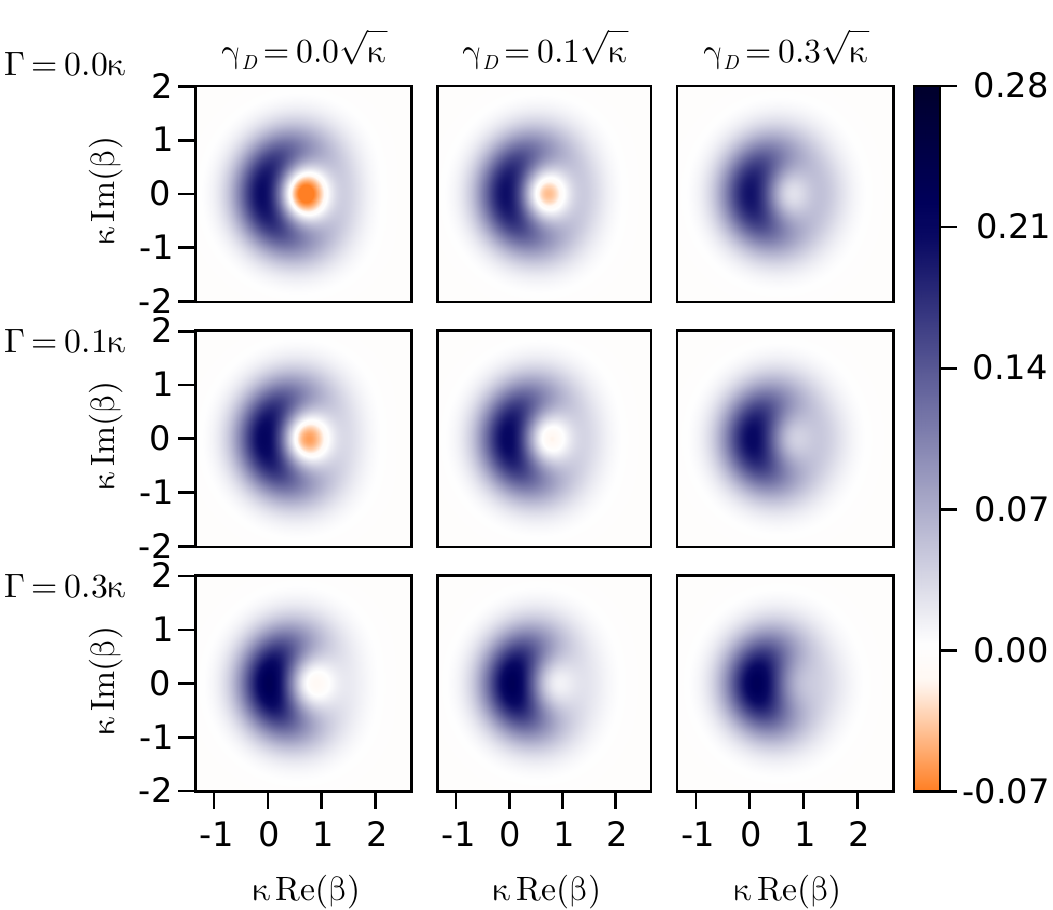}
    \caption{
        Wigner functions of $\rho_v$ at $\alpha = 0.5\sqrt{\kappa}$ for different $\Gamma$ and $\gamma_D$.
        The binning interval $(t_0, t_0 + \tau)$ is the same for all examples and was chosen to maximise the Wigner negativity at $\Gamma = 0 = \gamma_D$.
    }
    \label{fig:Dissipation}
\end{figure}

Figure~\ref{fig:Dissipation} shows the effect of these two noise-sources on the Wigner function of the output light for a single emitter and a coherent input with $\alpha = 0.5\sqrt{\kappa}$.
The Wigner functions remain negative for moderate noise, while becoming more and more Gaussian for larger values of $\Gamma$ and $\gamma_D$, respectively.
Most notably, however, both noise sources have qualitatively the same influence on $\rho_v$. The only noticeable difference is that the excitation transfer $\gamma_D$ suppresses the Wigner negativity slightly more than a similarly strong decay $\Gamma$.
This can be understood by noting that both noise sources result in Poissonian loss of excited state population, yet the decay into $|D\rangle$ also prohibits absorption and re-emission of subsequent photons into the mode $v(t)$.
However, as the binning interval is chosen such that approximately only one such event occurs, the difference to the spontaneous decay $\Gamma$ is minuscule.
Eventually, as the noise becomes sufficiently strong, the emitters becomes transparent to the incoming light and we find $\rho_v = |\sqrt{\tau}\alpha\rangle\langle\sqrt{\tau}\alpha|$, consequently.

\subsubsection{Multiple emitters}
Many of the single emitter results can be directly generalised to chains of multiple emitters.
However, in a chain of chirally coupled emitters the interaction~\eqref{eq:Hsys} between them and their collective decay through $D_L$ substantially impacts the dynamics of each individual emitter, making it impossible for them to emit simultaneously into the cavity at peak rates.
Consequently, $\rho_v$ becomes even more sensitive to the choice of the binning interval as some of the competing effects become enhanced compared to the single emitter case.
This section elucidates the key differences between the single emitter setup and the emitter chain and discusses the impact of chiral waveguide-mediated emitter interactions.
\begin{figure*}[t]
    \centering
    \includegraphics[width=\textwidth]{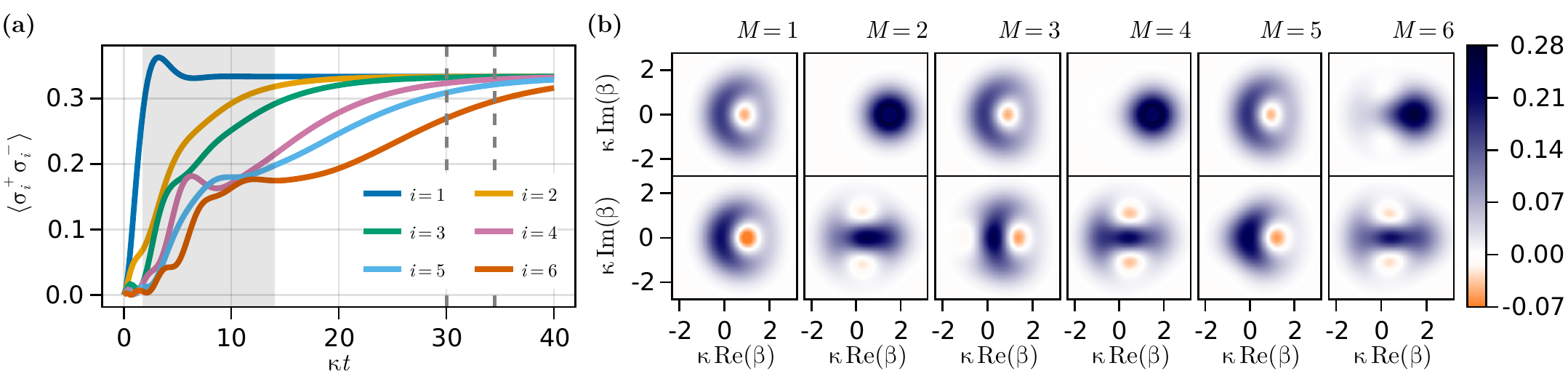}
    \caption{
        \textbf{(a)} Excited state population of the $i$-th emitter in a six emitter chain at $\alpha = 0.5 \sqrt{\kappa}$.
        The grey shaded region indicates the region which generally provide good bins for large Wigner negativities.
        The vertical dashed lines indicate the bin for top row of figure panel \textbf{(b)}, which shows the alternating pattern in the Wigner function, when binning in the emitters' steady state.
        \textbf{(b)} bottom: Wigner functions for one to six emitters.
        The binning interval was numerically optimised for each set of emitters and lie within the grey shaded region in figure panel (a).
    }
    \label{fig:chain}
\end{figure*}

In the short binning limit $\kappa \tau \ll 1$, the cavity state $\rho_v$ generated by $M$-emitters is a $M+1$-state mixture of $M$-photon added coherent states~\cite{Appendix}
\begin{align}
    \notag
    &\mathcal{D}^\dagger(\sqrt{\tau}\alpha) \rho_v
    \mathcal{D}(\sqrt{\tau}\alpha) \\
    \notag
    &\quad= \sum_{n=0}^M \sum_{m=0}^M
    \left\langle
        \big(\sqrt{\kappa \tau} \sigma_\mathrm{chain}^+\big)^{n}
        \big(\sqrt{\kappa \tau} \sigma_\mathrm{chain}^-\big)^{m}
    \right\rangle\\
    &\quad\quad\times
    \sum_{k=0}^{\min(n,m)} (-1)^k
    \frac{%
        |m-k \rangle\langle n-k|
    }{%
        k!\sqrt{(n-k)!}\sqrt{(m-k)!}
    },
    \label{eq:DensityMAtoms}
\end{align}
as each emitter contributes up to one photon to the cavity.
Yet, the Wigner function for the short bin density matrix~\eqref{eq:DensityMAtoms} will again be positive as the $n$-photon components are suppressed by at least $\sqrt{\kappa \tau}\,^n$.
Hence, sufficiently broad time bins $\kappa \tau \sim 1$ are required to obtain non-classical states $\rho_v$ like in the single emitter case.

Numerically, we find that a chain of emitters provides the largest Wigner negativities when $\tau$ is of the order of one Rabi cycle $1/\sqrt{\kappa}\alpha$ and for moderate driving strengths $\alpha \lesssim \sqrt{\kappa}$.
Due to the emitter interactions $H_\mathrm{sys}$, the excited state dynamics for emitters at the chain's end differ significantly from those of the first emitter.
Consequently, we can no longer choose the interval $(t_0, t_0 + \tau)$ such that it includes centers on Rabi peaks for all emitters' populations.
On the other hand, we again find that binning in the steady state inhibits Wigner negativity, and $\rho_v$ even reduces to a simple coherent state $|\sqrt{\tau}\alpha\rangle\langle\sqrt{\tau}\alpha|$, when the number of emitters is even, as we will explain in the following sections.
Hence, we find the largest Wigner negativities when the bin $(t_0, t_0 + \tau)$ starts at the onset of the excited state dynamics of the last emitter in the chain.
These results are exemplified in Figure~\ref{fig:chain}, where we show the dynamics of the emitters' excited state populations, panel~(a), and the Wigner functions of $\rho_v$,  panel~(b), for chains of up to six emitters.

The Wigner functions exhibit alternating features depending on whether of the number of emitters is even or odd, which becomes more prominent the farther the binning interval reaches into the steady state region of the excited state dynamics, Figure~\ref{fig:chain}(b) top.
This behaviour is well explained by the Bethe-state solutions for  propagating photons in a chiral emitter-chain~\cite{yudson1985,mahmoodian2020}, where the eigenstates of the full photon-emitter system are classified as scattering states and $n$-photon bound-states.
When scattering on a single emitter, each Bethe-state acquires an energy $E$ dependent phase
\begin{equation}
    t_{E,n} = \frac{E - i \kappa n^2 / 2}{E + i \kappa n^2 / 2},
\end{equation}
with $n = 1$ for the scattering states.
For time bins in the steady state regime we may ignore the ramp up process of the incoming light at $t = 0$ and approximate the incoming light by resonant plane waves.
Consequently, the light field primarily overlaps with the $E = 0$ Bethe-states and, after scattering at one emitter, every Bethe-state obtains a phase factor $-1$.
This then alters the photon state depending on the number of Bethe-states involved in the eigenstate decomposition.
For example, a two-photon state is decomposed into a product of two scattering states plus a single two-photon bound state, so that only the bound state picks up the $-1$-phase.
Consequently, the phases obtained by scattering at an even number of emitters $M$ in the steady state cancel each other, restoring the initial photon state, while odd $M$ change the photonic state.
More precisely, chains with even $M$ produce coherent output $\rho_v = |\sqrt{\tau}\alpha\rangle\langle\sqrt{\tau}\alpha|$, as long as the bin $(t_0, t_0 + \tau)$ overlaps with the steady state region of each emitter.

The bottom row of Figure~\ref{fig:chain}~(b) shows the Wigner function for different $M$ for early time bins outside the steady state regime, which were chosen separately for each $M$ to maximise negativity.
Qualitatively, we obtain similar results as above except that negative features now also occur for even $M$. 
The alternating pattern of the negative features again follow from the parity of the phase factor $(-1)^M$ obtained after scattering on $M$ emitters.
The incoming light field may now be considered as resonant planes waves plus a correction due to the ramp up at $t = 0$, which allows for higher Wigner negativities than in the steady state regime.
In coordinate space, however, the transfer matrix~$t_{E,n}$ acts as a convolution with kernel $\delta(x) - \kappa n^2 e^{-\kappa n^2 x/2}\theta(x)$.
Therefore, the spatial profile of the correction broadens after each subsequent emitter, reducing its overlap with the projection mode $v(t)$.
While we overall benefit from using multiple emitters in the creation of non-classical $\rho_v$, the general structure of the Wigner functions are already known after studying two emitters and the Wigner negativity eventually settles to the respective steady state value.
With respect to maximizing the Wigner negativity, we find no significant benefit in using more than $M=4$ emitters.

\subsection{Application example: Interferometry}
\begin{figure*}[t]
    \centering
    \includegraphics[width=\textwidth]{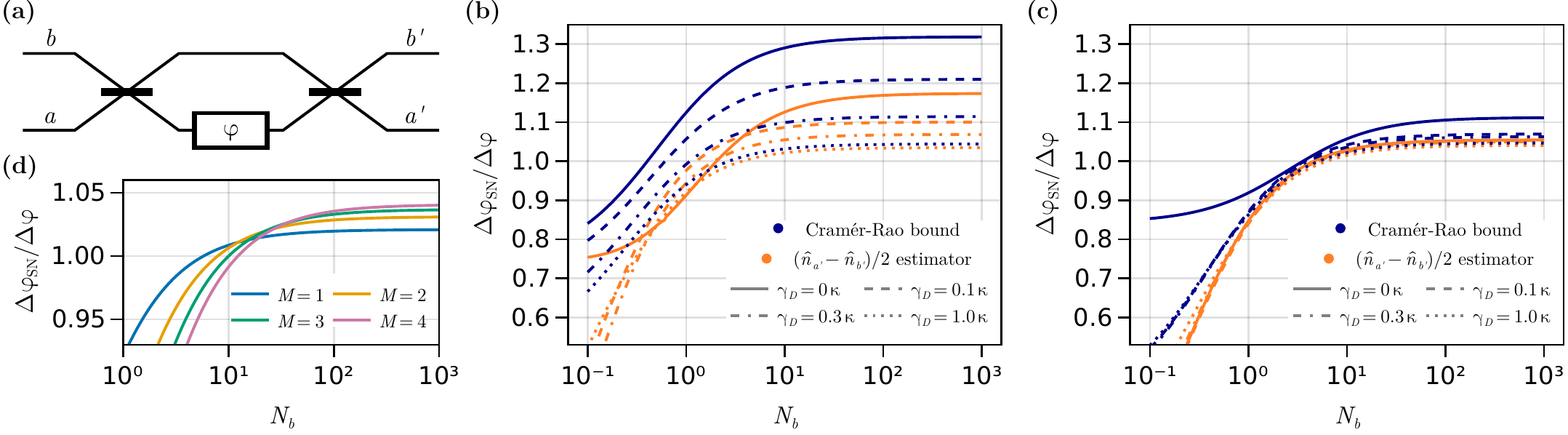}
    \caption{
        \textbf{(a)} Sketch of the Mach-Zehnder experiment.
        Light on the two input ports $a$ and $b$ interferes at the beam splitters and measurements on the output ports $a'$ and $b'$ are used to estimate the unknown phase $\varphi$.
        \textbf{(b)}, \textbf{(c)} We use $\rho_v$, generated from one and two emitters respectively, at the input port $a$ and a coherent state $|\sqrt{N_b}\rangle$ at port $b$ to estimate $\varphi$.
        The figures show the improvement of the $\Delta\varphi$-sensitivity and Cram\'er-Rao bound compared to shot noise $\Delta \varphi_\mathrm{SN} = 1/\sqrt{N_a + N_b}$ at multiple values $\gamma_D$.
        For any finite $\gamma_D$ the optimal driving strength $\alpha$ and bin $(t_0, t_0 + \tau)$ depend little on $N_b$.
        However, this is not the case for a single emitter at $\gamma_D = 0$ and we choose to optimise at $N_b = 100$, after which the sensitivities typically approach their asymptotic values.
        \textbf{(d)} At high decay $\gamma_D = 2\kappa$ multiple emitters provide more noise-resilient $\Delta\varphi$-sensitivity improvements.
    }
    \label{fig:MachZehnder}
\end{figure*}
The creation of the non-classical state of light $\rho_v$ relies purely on the interaction of the classical state $|\alpha\rangle$ with the emitters and is thus deterministic, making $\rho_v$ of interest for applications where no post-selection is desired.
However, we have to keep in mind that the scattering of the coherent input on the emitters produces light in multiple orthogonal modes.
Our approach cannot describe the entire photon state at time~$t_0 + \tau$, nor do we account for all photons within the spatial bin $(t_0, t_0 + \tau)$.
This lack of information implies that $\rho_v$ is not directly accessible in potential applications.
In this section, we remedy this shortcoming and show an example how $\rho_v$ may be used in quantum metrology experiments.

The cavity state $\rho_v$ does not account for all photons in the time bin $(t_0, t_0 + \tau)$, as we only considered a single temporal mode $v(t)$.
Nevertheless, we may use the light emitted from the emitter-chain in places where we want to use $\rho_v$ as a resource, as long as mode-mixing does not occur, since we can selectively measure the light in mode $v(t)$ via homodyne detection.
This is the case in experiments consisting of linear optical devices.
Under this constraint, we discuss a possible application in quantum metrology where the combination of $\rho_v$ and a coherent state as the two inputs to a Mach-Zehnder interferometer can outperform the standard quantum limit of interferometry.

We consider the setup depicted in Figure~\ref{fig:MachZehnder}~(a), where an unknown phase~$\varphi$ in one of the interferometer arms shall be determined.
In the standard quantum limit of interferometry~\cite{demkowicz2015, wiseman2009} coherent light is used, and $\varphi$ is estimated by the intensity-difference at the output ports
\begin{equation}
    J_z = \frac{\hat{n}_{a'} - \hat{n}_{b'}}{2}.
    \label{eq:Estimator}
\end{equation}
The best achievable precision with this estimator is
\begin{equation}
    \Delta\varphi = \min_\varphi
        \frac{\Delta J_z}{%
            \left|\frac{\partial \langle J_z \rangle}{\partial \varphi}\right|%
        }
    \label{eq:Sensitivity}
\end{equation}
and yields shot-noise precision $\Delta\varphi_\mathrm{SN} = 1 / \sqrt{N}$ in the standard quantum limit, where $N$ is the total number of photons in both input ports.
The precision of a given estimator, however, may be significantly improved when non-classical states are used as input ports~\cite{caves1981,Yurke1986}.

The single-emitter-$\rho_v$ achieves such an improvement in $\Delta\varphi$ when interfering with a sufficiently strong coherent state $|\sqrt{N_b}\rangle$.
Figure~\ref{fig:MachZehnder}~(b) shows the obtained precision, which consistently beats the shot noise limit $\Delta\varphi_\mathrm{SN} = 1 / \sqrt{N_a + N_b}$ already for moderate photon numbers $N_b > 10$.
For a fair comparison, we calculate the shot-noise with no emitter present, i.e., $N_a = \tau |\alpha|^2$, where all photons in the time bin $(t_0, t_0 + \tau)$ can contribute to the measurement.
In the asymptotic limit and with weak decay $\gamma_D = 0.1\kappa$, we numerically find an improvement of about $10\,\%$ with the estimator $J_z$.
The Cram\'er-Rao bound, which bounds the highest obtainable precision with any estimator, reveals the possibility to more than double the sensitivity improvement to $21\,\%$.
The auxiliary state $\rho_v$ should be compared to squeezed vacuum states, which are the typically used to improve the sensitivity.
A squeezed state with the same mean photon number as $\rho_v$ provides an improvement of $30\,\%$.
While squeezed states outperform $\rho_v$, the simplicity of creating $\rho_v$ still renders it a promising alternative.

As we have seen in the last section, we generally should not expect any major benefits in using more than 2 emitters to generate nonclassicality in $\rho_v$ in absence of noise.
This is verified by the results in Figure~\ref{fig:MachZehnder}~(c), which shows the Cram\'er-Rao bound and $\Delta\varphi$ compared to shot noise precision for $\rho_v$ generated with two emitters.
While the sensitivity-improvement falls well below the single-emitter-$\rho_v$ results, it becomes far more resilient to dissipation.
We explain this as follows: Once the decay rate becomes the dominant timescale our previous bound-state analysis is no longer valid, and the light after an emitter differs only slightly from the incoming state as already shown in Figure~\ref{fig:Wigner}.
These small corrections, however, are amplified by scattering multiple times. 
Thus adding more emitters in the high dissipation regime can enhance the non-classical features and thus provide a robust sensitivity improvement.
As can be seen in Figure~\ref{fig:MachZehnder}~(d), even for large decay rates $\gamma_D = 2\kappa$ we can still beat shot noise with the standard estimator~\eqref{eq:Estimator} by almost $5\,\%$.

\section{Conclusion}
In summary, we have used the input-output theory to show that coherently driven chains of quantum emitters generate light modes with non-classical number statistics.
However, with the current setup of a constant driving field and the flat mode $v(t)$, the obtained non-classicality depends strongly on the chosen binning interval, especially since the emission in the steady-state regime shows weak features of non-classicality.
We saw that the non-classicality originates from the emission of a single energy quanta after the decay of one of the emitters.
Therefore, we propose that an individual emitter, periodically driven between $|G\rangle$ and $|W\rangle$, will produce stronger non-classicalities, while also being less sensitive to the exact binning parameters, as long as the binning width $\tau$ is commensurable to the emitter's period.

A periodic evolution of the emitter state is only possible for time-dependent driving strengths $\alpha(t)$.
While our formalism allows the study of non-constant $\alpha(t)$ without any modification, we expect many of the observed effects to change.
For example, we expect that it becomes beneficial under time-dependent driving to use more than two superatoms for the generation of non-classical light, as the alternating pattern in the Wigner functions was only observed due to the large overlap of the driving field with the $E = 0$ Bethe-states.
This will not be the case, however, when $\alpha(t)$ changes significantly on timescales $1/\kappa$.
At the same time, we also expect that the temporal profile of the output mode $v(t)$ in the interval $(t_0, t_0 + \tau)$ should also change in time to better suit the non-constant drive.
The optimal profile of $v(t)$, however, likely has to be determined by numerical optimisation.

Since chiral waveguides are implemented in many systems, like superconducting circuits~\cite{hoi2012,chapman2017}, photonic crystal waveguides~\cite{arcari2014,sollner2015,xiao2021}, or Rydberg superatoms~\cite{paris2017}, and since $\rho_v$ is directly accessible in linear quantum optical systems and via homodyne detection, we identify the proposed setup as a promising candidate for the creation of non-classical states of light.

\begin{acknowledgments}
\section*{Acknowledgements and Funding}
This work has received funding from the European Union’s Horizon 2020 programme
under the ERC consolidator grants SIRPOL (grant no. 681208) and RYD-QNLO (grant
no. 771417), the ErBeStA project (grant no. 800942),  the Deutsche Forschungsgemeinschaft
(DFG) under SPP 1929 GiRyd project BU 2247/4-1, and the
Carlsberg Foundation through the Semper Ardens Research Project QCooL.
\end{acknowledgments}

\bibliographystyle{apsrev4-1}
\bibliography{references}

\onecolumngrid
\appendix
\section{Short bin density matrix}
\label{app:MasterEquation}
The density matrix $\rho_v$ of the photons in mode $v$ can be found by explicitly calculating each matrix element as
\begin{equation}
    (\rho_v)_{m,n} = \Big\langle |n\rangle\langle m| \Big\rangle
    = \frac{1}{\sqrt{n!m!}}
    \big\langle :(b_v^\dagger)^n e^{-b_v^\dagger b_v} b_v^m : \big\rangle.
    \label{eq:DensityGeneral}
\end{equation}
Here, $:f(b_v^\dagger, b_v):$ denotes the normal ordering of $f(b_v^\dagger, b_v)$.
Since
\begin{equation}
    b_v = \frac{1}{\sqrt{\tau}} \int_{t_0}^{t_0+\tau} dt
    \big( \alpha + \sqrt{\kappa} \sigma_\mathrm{chain}^-(t)\big),
\end{equation}
the relation~\eqref{eq:DensityGeneral} for $\rho_v$ generally yields out-of-time-ordered correlation functions and thus determining $\rho_v$ is impractical in most situations.
However, for $\kappa \tau \ll 1$ we may approximate $b_v \approx \sqrt{\tau}\big(\alpha + \sqrt{\kappa} \sigma_\mathrm{chain}^-(t_0)\big)$, and thus calculate $\rho_v$ from the emitters' density matrix at time $t_0$.

Assuming $M$ chiral emitter we can expand the expectation value in~\eqref{eq:DensityGeneral} as
\begin{align}
    \notag
    :(b_v^\dagger)^n e^{-b_v^\dagger b_v} b_v^m :
    &\ = \sqrt{\tau}^{n+m}\sum_{k=0}^\infty \frac{(-\tau)^k}{k!}
    \big(\alpha^* + \sqrt{\kappa} \sigma_\mathrm{chain}^+\big)^{n+k}
    \big(\alpha   + \sqrt{\kappa} \sigma_\mathrm{chain}^-\big)^{m+k} \\
    \notag
    &\ = \sqrt{\tau}^{n+m}
    \sum_{\tilde{n}=0}^M (\alpha^*)^{n-\tilde{n}} (\sqrt{\kappa} \sigma_\mathrm{chain}^+)^{\tilde{n}}
    \sum_{\tilde{m}=0}^M  \alpha^{m-\tilde{m}} (\sqrt{\kappa} \sigma_\mathrm{chain}^-)^{\tilde{m}}\\
    &\ \qquad\times
    \sum_{k=0}^\infty
    \frac{(-\tau |\alpha|^2)^k}{k!} \binom{n+k}{\tilde{n}}\binom{m+k}{\tilde{m}}.
\end{align}
Due to the coherent drive it is expected that $\rho_v$ possesses large overlap with $|\sqrt{\tau}\alpha\rangle$.
Therefore, we extract a factor $e^{-\tau|\alpha|^2}$ from the $k$-summation by inserting $1 = e^{-\tau |\alpha|^2} e^{\tau|\alpha|^2}$, expanding the positive exponential and thereon collecting all terms of equal power in $(\tau|\alpha|^2)^k$, resulting in
\begin{align}
    \notag
    :(b_v^\dagger)^n e^{-b_v^\dagger b_v} b_v^m :
    &\ =
    (\sqrt{\tau}\alpha^*)^n (\sqrt{\tau}\alpha)^m e^{-\tau|\alpha|^2}
    \sum_{\tilde{n}=0}^M \frac{1}{\tilde{n}!} \left(\frac{\sqrt{\kappa} \sigma_\mathrm{chain}^+}{\alpha^*}\right)^{\tilde{n}}
    \sum_{\tilde{m}=0}^M \frac{1}{\tilde{m}!} \left(\frac{\sqrt{\kappa} \sigma_\mathrm{chain}^-}{\alpha}\right)^{\tilde{m}}\\
    \notag
    &\quad\qquad\times
    \sum_{k=0}^\infty
    \frac{(\tau |\alpha|^2)^k}{k!} \sum_{i=0}^k (-1)^i \binom{k}{i} (n+i)^{\ff{\tilde{n}}} (m+i)^{\ff{\tilde{m}}} \\
    \notag
    &\ = \sqrt{n!m!}
    \langle m|\sqrt{\tau}\alpha\rangle\langle\sqrt{\tau}\alpha| n\rangle
    \sum_{\tilde{n}=0}^M \frac{1}{\tilde{n}!} \left(\frac{\sqrt{\kappa} \sigma_\mathrm{chain}^+}{\alpha^*}\right)^{\tilde{n}}
    \sum_{\tilde{m}=0}^M \frac{1}{\tilde{m}!} \left(\frac{\sqrt{\kappa} \sigma_\mathrm{chain}^-}{\alpha}\right)^{\tilde{m}}\\
    &\quad\qquad\times
    \sum_{k=0}^\infty
    \frac{(- \tau |\alpha|^2)^k}{k!} \fd^k\Big|_{x=0} (n+x)^{\ff{\tilde{n}}} (m+x)^{\ff{\tilde{m}}}.
    \label{eq:DensityIntermediate}
\end{align}
Here, we introduced $x^{\ff{n}} = x(x-1)\dots(x-n+1)$ the falling factorial, and the forward difference operator $\fd$, defined as $\fd f(x) = f(x+1) - f(x)$.
For the simplification in the second step we used
\begin{equation}
    (-1)^k\fd^k f(x) = \sum_{i=0}^k (-1)^i \binom{k}{i} f(x+i).
\end{equation}
Next, we eliminate $n$ and $m$ from~\eqref{eq:DensityIntermediate} by using the number operator $n|n\rangle = b^\dagger b |n\rangle$ and find the density matrix
\begin{align}
    \notag
    &\rho_v = \sum_{\tilde{n}=0}^M \sum_{\tilde{m}=0}^M
    \frac{1}{\tilde{n}!} \frac{1}{\tilde{m}!}
    \left\langle
    \left(\frac{\sqrt{\kappa} \sigma_\mathrm{chain}^+}{\alpha^*}\right)^{\tilde{n}}
    \left(\frac{\sqrt{\kappa} \sigma_\mathrm{chain}^-}{\alpha  }\right)^{\tilde{m}}
    \right\rangle\\
    \notag
    &\quad\qquad\times
    \sum_{k=0}^{\infty} \frac{(- \tau |\alpha|^2)^k}{k!} \fd^k\Big|_{x=0}
    (b^\dagger b + x)^{\ff{\tilde{m}}} |\sqrt{\tau}\alpha\rangle\langle\sqrt{\tau}\alpha| (b^\dagger b + x)^{\ff{\tilde{n}}}.
    \label{eq:DensityPre}
\end{align}
It already is evident that $\rho_v$ is generated from the coherent state $|\sqrt{\tau}\alpha\rangle\langle\sqrt{\tau}\alpha|$ by application of an operator which is a function in $b^\dagger b$.
Next, we show that this operator only adds up to $M$ photons to $\rho_v$ and find $\rho_v$ in a Fock-state-basis.

The algebra of finite differences with falling factorials posses many similarities to the derivatives of monomials, e.g.~$\fd x^{\ff{n}} = n x^{\ff{n-1}}$, and one finds the generalised product rule $\fd fg = (\fd f)g + f(\fd g) + (\fd f)(\fd g)$.
Thus it is evident that~\eqref{eq:DensityPre} will only consist of falling factorials of the number operator, which directly translate into normal ordered powers $(b^\dagger b)^{\ff{n}} = :(b^\dagger b)^n:$.
Therefore, $\rho_v$ in~\eqref{eq:DensityPre} is invariant under the set of replacements
\begin{align}
    \fd\Big|_{x=0} \ &\mapsto\quad
    (\partial_x + \partial_y + \partial_x\partial_y)\Big|_{x=0=y} \\
    (b^\dagger b + x)^{\ff{\tilde{m}}} |\sqrt{\tau} \alpha\rangle \ &\mapsto\quad
    :(b^\dagger b + x)^{\tilde{m}}: |\sqrt{\tau} \alpha\rangle
    = (\sqrt{\tau}\alpha b^\dagger + x)^{\tilde{m}} |\sqrt{\tau} \alpha\rangle
    = \mathcal{D}(\sqrt{\tau}\alpha)
    (\sqrt{\tau}\alpha b^\dagger + \tau|\alpha|^2 + x)^{\tilde{m}} |0\rangle
    \label{eq:ReplaceLeft}\\
    \langle\sqrt{\tau} \alpha| (b^\dagger b + x)^{\ff{\tilde{n}}} \ &\mapsto\quad
    \langle\sqrt{\tau} \alpha| :(b^\dagger b + y)^{\tilde{n}}:\  
    = \langle\sqrt{\tau} \alpha| (\sqrt{\tau}\alpha^* b + y)^{\tilde{n}}
    = \langle0| (\sqrt{\tau}\alpha^* b + \tau|\alpha|^2 + y)^{\tilde{n}}
    \mathcal{D}^\dagger(\sqrt{\tau}\alpha).
    \label{eq:ReplaceRight}
\end{align}
We now perform the $k$-summation, which yields two translation operators $T_{x,y}(-\tau|\alpha|^2)$ for $x, y$ and the operator $\exp(-\tau|\alpha|^2 \partial_x\partial_y)$.
The translation operators cancel the $\tau|\alpha|^2$ terms in \eqref{eq:ReplaceLeft} and~\eqref{eq:ReplaceRight}. After rescaling $x$, $y$, we end up with the density matrix
\begin{align}
    \notag
    \mathcal{D}^\dagger(\sqrt{\tau}\alpha) \rho_v
    \mathcal{D}(\sqrt{\tau}\alpha)
    &= \sum_{\tilde{n}=0}^M \sum_{\tilde{m}=0}^M
    \frac{1}{\tilde{n}!}\frac{1}{\tilde{m}!}
    \left\langle
        \big(\sqrt{\kappa \tau} \sigma_\mathrm{chain}^+\big)^{\tilde{n}}
        \big(\sqrt{\kappa \tau} \sigma_\mathrm{chain}^-\big)^{\tilde{m}}
    \right\rangle
    e^{-\partial_x\partial_y}\Big|_{x=0=y} 
    (b^\dagger + x)^{\tilde{m}} |0\rangle\langle0| (b + y)^{\tilde{n}}\\
    &= \sum_{\tilde{n}=0}^M \sum_{\tilde{m}=0}^M
    \left\langle
        \big(\sqrt{\kappa \tau} \sigma_\mathrm{chain}^+\big)^{\tilde{n}}
        \big(\sqrt{\kappa \tau} \sigma_\mathrm{chain}^-\big)^{\tilde{m}}
    \right\rangle
    \sum_{k=0}^{\min(\tilde{n},\tilde{m})} (-1)^k
    \frac{%
        |\tilde{m}-k \rangle\langle \tilde{n}-k|
    }{%
        k!\sqrt{(\tilde{n}-k)!}\sqrt{(\tilde{m}-k)!}
    }.
\end{align}
Here, the right hand side is spanned by the truncated Fock space $\{|0\rangle,\dots,|M\rangle\}$, which is to say that $\rho_v$ generally is a $(M+1)$-state mixture of $M$-photon added coherent states in the short bin limit $\kappa \tau \ll 1$.

\end{document}